\documentclass[12pt]{article}
\usepackage{graphicx}
\usepackage{epstopdf}
\usepackage{caption}
\usepackage{subcaption}
\usepackage{subfig}

\usepackage{epsfig,amsmath,amssymb,verbatim,mathrsfs}

\def\beq{\begin{eqnarray}}
\def\eeq{\end{eqnarray}}
\def\bea{\begin{eqnarray}}
\def\eea{\end{eqnarray}}

\newcommand{\f}{\mathcal{F}}

\renewcommand{\thefootnote}{\roman{footnote}}

\begin{document}

\setlength{\baselineskip}{0.2in}


\begin{titlepage}
\noindent
\flushright{April 2016}
\vspace{0.2cm}

\begin{center}
  \begin{Large}
    \begin{bf}
A Critical Analysis of One-Loop Neutrino Mass Models with Minimal Dark Matter

     \end{bf}
  \end{Large}
\end{center}


\begin{center}

\begin{bf}
{Amine~Ahriche,$^{1,2,}$\footnote{aahriche@ictp.it}
Kristian~L.~McDonald,$^{3,}$\footnote{kristian.mcdonald@sydney.edu.au} 
Salah~Nasri$^{4,}$\footnote{snasri@uaeu.ac.ae}} and Ivica~Picek$^{5,}$\footnote{picek@phy.hr}\\
\end{bf}
%

 \begin{it}
$^1$ Laboratory of Mathematical and Sub-Atomic Physics (LPMPS), University of
Constantine I, DZ-25000 Constantine, Algeria\\
\vspace{0.1cm}
$^2$ The Abdus Salam International Centre for Theoretical Physics, Strada Costiera 11, I-34014, Trieste, Italy\\
\vspace{0.1cm}
$^3$ ARC Centre of Excellence for Particle Physics at the Terascale,\\
School of Physics, The University of Sydney, NSW 2006, Australia\\
\vspace{0.1cm}
$^4$ Physics Department, UAE University, POB 17551, Al Ain, United Arab Emirates\\
\vspace{0.1cm} 
$^5$ Department of Physics, Faculty of Science,  University of Zagreb, P.O.B. 331, \\HR-10002 Zagreb, Croatia \vspace{0.3cm}
\end{it}

\vspace{0.5cm}

\end{center}


\begin{abstract}

A recent paper investigated minimal R$\nu$MDM models  with the type T1-iii and T3 one-loop topologies. However, the candidate most-minimal model does not possess an accidental symmetry - the scalar potential contains an explicit symmetry breaking term, rendering the dark matter unstable. We present two models that cure this problem.  However, we further show that \emph{all} of the proposed minimal one-loop R$\nu$MDM models  suffer from a second problem - an additional source of explicit $Z_2$ symmetry breaking in the Yukawa sector. We perform a more-general analysis to show that neutrino mass models using either the type T3 or type T1-iii one-loop topologies do not give viable minimal dark matter candidates. Consequently, one-loop models of neutrino mass with minimal dark matter do not appear possible. Thus, presently there remains a single known (three-loop) model of neutrino mass that gives stable dark matter without invoking any new symmetries.

\end{abstract}

\vspace{1cm}

\end{titlepage}
\renewcommand{\thefootnote}{\arabic{footnote}}
\setcounter{footnote}{0}
\setcounter{page}{1}


\vfill\eject


\section{Introduction\label{sec:introduction}}

The origin of neutrino mass and the particle physics properties of  dark matter (DM) constitute two important unsolved problems in particle physics research. While it is a logical possibility that these problems possess independent solutions, it is interesting to consider the alternative -  that they admit a common or unified solution. This was the approach advocated by Krauss, Nasri and Trodden~\cite{Krauss:2002px} and also by Ma~\cite{Ma:2006km}; both groups presented models in which neutrino mass appears as a radiative effect due to interactions with a $Z_2$-odd sector that contains a stable DM candidate. The former (latter) advocated a three-loop (one-loop) model of neutrino mass. More generally there has been a great deal of research in this area; for related early works see Ref.~\cite{Aoki:2008av}, while for more recent models see  e.g., Ref.~\cite{Hatanaka:2014tba} and  references therein.

In the Standard Model (SM) proton stability results from an accidental (baryon number) symmetry. It is natural to ask whether DM stability could similarly result from an accidental symmetry. This approach, dubbed Minimal DM~\cite{Cirelli:2005uq}, is well studied in the literature. In the context of the SM, it is well known that an accidentally-stable DM candidate arises if the SM is extended to include either a hypercharge-less quintuplet fermion multiplet, $\mathcal{F}\sim(1,5,0)$, or a septuplet scalar multiplet, $\phi\sim(1,7,0)$~\cite{Cirelli:2005uq}. Note that the Minimal DM framework does not hold for a scalar multiplet $\phi\sim(1,5,0)$, as the $Z_2$ symmetry is explicitly broken~\cite{Cirelli:2005uq}.

The notion of Minimal DM was first applied to radiative neutrino mass models in Ref.~\cite{Cai:2011qr}, the goal being to extend the SM with new particles that generate radiative neutrino mass while also giving an accidental symmetry to achieve a stable DM candidate (the model was dubbed R$\nu$MDM). Unfortunately it was subsequently shown that the model did not work, due to a symmetry-breaking term in the scalar potential~\cite{Kumericki:2012bf}. More recently a three-loop model of neutrino mass was proposed in which DM stability resulted from an accidental symmetry (without invoking any beyond-SM symmetries)~\cite{Ahriche:2015wha}. This appears to be the first viable model to achieve accidental DM in the context of a radiative neutrino mass model, the DM being a septuplet fermion, $\mathcal{F}\sim(1,7,0)$, in this instance. There also exists a three-loop model of neutrino mass~\cite{Culjak:2015qja} that employs both minimal DM candidates identified in Ref.~\cite{Cirelli:2005uq}.

Motivated by a recent study of one-loop models for neutrino mass with minimal DM~\cite{Cai:2016jrl}, we perform a general analysis of one-loop models. In particular, we show that neutrino mass models using either the type T3 or type T1-iii one-loop topologies~\cite{Bonnet:2012kz} do not give viable (i.e.~accidentally stable) minimal DM candidates, due to explicit breaking of the requisite symmetry. Furthermore, our results indicate that it is not  possible to obtain minimal DM by the use of one-loop neutrino mass models.

The layout of  this paper is as follows. In Section~\ref{sec:scalar_T1-iii} we demonstrate the presence of explicit $Z_2$ symmetry breaking in the scalar potential of the minimal one-loop model identified in Ref.~\cite{Cai:2016jrl}, presenting two new models that cure this problem. In Section~\ref{sec:T3} we perform a critical analysis of models with the type T3 one-loop topology. Similarly, we study models with the type T1-iii topology in detail in Section~\ref{sec:T1}. Conclusions are drawn in Section~\ref{sec:conc}.


\section{Symmetry Breaking in the Scalar Potential for Type T1-iii One-Loop Models\label{sec:scalar_T1-iii}}

\begin{figure}[ttt]
\begin{center}
        \includegraphics[width = 0.50\textwidth]{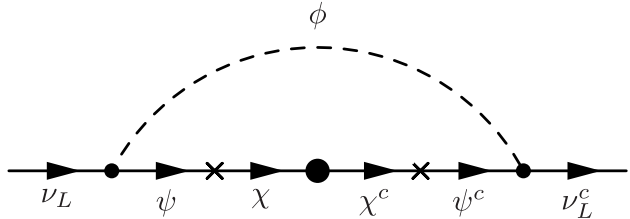}
\end{center}
\caption{One-loop diagram for neutrino mass with the type T1-iii topology. Crosses denote insertions of the SM Higgs vacuum value and the larger dot denotes a Majorana mass insertion for the real fermion $\chi$.}\label{fig:one_loop}
\end{figure}

A recent  work has reconsidered the R$\nu$MDM approach, attempting to find one-loop neutrino mass models with accidentally stable DM candidates~\cite{Cai:2016jrl}. Three models were identified as candidates; two employing the so-called T1-iii one-loop topology~\cite{Bonnet:2012kz}, with the beyond-SM particle content being (see Table~1 in Ref.~\cite{Cai:2016jrl})\footnote{See the Note Added at the end of the paper. Also, note that our hypercharge normalization differs by a factor of 2.}
\bea
\mathrm{Model\ A}: \quad \chi\sim(1,7,0),\quad \psi\sim(1,6,1),\quad \bar{\psi}\sim(1,6,-1),\quad \phi\sim(1,5,0),\label{modela}\\
\mathrm{Model\ B}:\quad\chi\sim(1,5,0),\quad \psi\sim(1,4,1),\quad \bar{\psi}\sim(1,4,-1),\quad \phi\sim(1,5,0),\label{modelb}
\eea
where $\chi$, $\psi$, and $\bar{\psi}$ are fermions while $\phi$ is a scalar multiplet (the type T1-iii one-loop diagram for these models is shown in Figure~\ref{fig:one_loop}). The models are purportedly invariant under an  accidental $Z_2$ symmetry, where $\chi$, $\psi$, $\bar{\psi}$ and $\phi$ are $Z_2$-odd, while the SM fields are $Z_2$-even; it is clear from Figure~\ref{fig:one_loop} that this is an accidental symmetry of the loop diagram. One further model, employing the so-called T3 one-loop topology was also proposed. Ref.~\cite{Cai:2016jrl} then performed a detailed study of the T1-iii one-loop model with particle content in Eq.~\eqref{modelb}, namely Model B.

With regard to Models A and B in Eqs.~\eqref{modela} and \eqref{modelb}, we note that  the most-general Lagrangian obtained by adding $\phi\sim(1,5,0)$ to the SM contains the term $\mu\phi^3$, which explicitly breaks any $Z_2$ symmetry under which $\phi$ is odd-valued. This point is understood by the absence of a scalar quintuplet Minimal DM candidate in Ref.~\cite{Cirelli:2005uq}. For completeness, however, we note that, after writing the scalar quintuplet in symmetric-tensor notation as $\phi_{abcd}$, where the $SU(2)$ indices take values $a,b,..=1,2$, the cubic term  $\mu\,\phi_{abcd}\, \phi_{efgh}\, \epsilon^{cg}\,\epsilon^{dh}\,( \phi^*)^{abef}$ appears in the most-general scalar potential (here $\mu$ denotes the coupling). This conclusion holds when additional fields are added to the model. Thus, Model B, defined by Eq.~\eqref{modelb} and studied in detail in Ref.~\cite{Cai:2016jrl}, contains an explicit source of $Z_2$ symmetry breaking and the DM candidate is unstable.

Here we wish to emphasize that a viable one-loop model for Minimal DM and radiative neutrino mass via the T1-iii topology  is obtained  if one modifies Model B by promoting the field content as follows:\footnote{Note that the term $\mu\phi^3$ is not allowed for a septuplet scalar, unlike the quintuplet case of Model B.}
\bea
\mathrm{Model\ C}: \quad\chi\sim(1,5,0),\quad \psi\sim(1,6,1),\quad \bar{\psi}\sim(1,6,-1),\quad \phi\sim(1,7,0).\label{modelC}
\eea
This model appears particularly interesting as it contains two DM candidates, namely the quintuplet fermion, $\chi\sim(1,5,0)$, and the septuplet scalar, $\phi\sim(1,7,0)$, both of which were identified as Minimal DM candidates in Ref.~\cite{Cirelli:2005uq}. Depending on the mass ordering of $\chi$ and $\phi$, it appears that either fermionic or scalar DM is possible (or possibly both in a near degenerate case). Importantly, the model does not contain the cubic term $\phi^3$ in the scalar potential.

With regards to Model A, we suspect that Table~1 in Ref.~\cite{Cai:2016jrl} (i.e.~Eqs.~\eqref{modela} and \eqref{modelb} above) contains a minor typographical error, in which the scalar $\phi$ should instead be a septuplet, $\phi\sim(1,5,0)\rightarrow \phi\sim(1,7,0)$. If this is correct, the scalar potential for Model A preserves the accidental symmetry of the loop diagram, though, unlike Model B, this model was not studied in detail. It is important to emphasize, however, that this model contains two DM candidates, both a fermionic DM candidate, in the form of the septuplet $\chi\sim(1,7,0)$~\cite{Ahriche:2015wha} and a scalar DM candidate, in the form of $\phi\sim(1,7,0)$. According to the criterion of minimality employed in Ref.~\cite{Cai:2016jrl}, it appears that Model C in Eq.~\eqref{modelC} would be considered more minimal, due to the smaller $SU(2)$ representations involved.

We also note that a related model is obtained by promoting the fermions $\psi$ in Model A to octuplets: 
\bea
\mathrm{Model\ D}: \quad \chi\sim(1,7,0),\quad \psi\sim(1,8,1),\quad \bar{\psi}\sim(1,8,-1),\quad \phi\sim(1,7,0).
\eea
This model may also be an interesting variant as it contains two distinct DM candidates, namely the fermion $\chi\sim(1,7,0)$ and the scalar $\phi\sim(1,7,0)$, though Model C would be considered more minimal.

In summary, we find the following minimal models of one-loop neutrino mass via the T1-iii topology in which the accidental symmetry of the one-loop diagram is preserved by the scalar potential:
\bea
\mathrm{Model\ I}: \quad \chi\sim(1,5,0),\quad \psi\sim(1,6,1),\quad \bar{\psi}\sim(1,6,-1),\quad \phi\sim(1,7,0),\nonumber\\
\mathrm{Model\ II}: \quad\chi\sim(1,7,0),\quad \psi\sim(1,6,1),\quad \bar{\psi}\sim(1,6,-1),\quad \phi\sim(1,7,0).\nonumber\\
\mathrm{Model\ III}: \quad \chi\sim(1,7,0),\quad \psi\sim(1,8,1),\quad \bar{\psi}\sim(1,8,-1),\quad \phi\sim(1,7,0),\label{eq:modelsI}
\eea
where Model II appears in Ref.~\cite{Cai:2016jrl} and Models I and III are new. The new Model I contains both Minimal DM candidates identified in Ref.~\cite{Cirelli:2005uq}, namely the fermion $\chi\sim(1,5,0)$ and the scalar $\phi\sim(1,7,0)$, and could give  either fermionic or scalar DM. None of these models possess the cubic term $\phi^3$, so the accidental symmetry of the loop diagram is preserved by the scalar potential. We note that, due to the fact that the quintuplet and septuplet DM candidates must have $\mathcal{O}(10)$~TeV masses~\cite{Farina:2013mla}, and one of these states must be the lightest exotic in the spectrum, the new multiplets would be well beyond the reach of the LHC; prospects for testing such models thus appear poor.\footnote{This differs from the three-loop model of Ref.~\cite{Ahriche:2015wha}, which contains an electrically-charged singlet scalar that is neutral under the accidental symmetry and can thus be lighter than the DM and within reach of the LHC. Similarly Ref.~\cite{Culjak:2015qja} contains a second Higgs doublet that can remain at the TeV scale.} However, searches for galactic gamma-ray signals can provide promising ways to test such models - in fact, subject to unknown dependencies on the cuspiness of the DM halo, one can already rule out the quintuplet and septuplet  DM candidate in regions of parameter space~\cite{Cirelli:2015bda}. Thus, the models are timely as they predict signals that are directly relevant for current and next-generation galactic  gamma-ray searches.

Given the apparent interest of these models, it is worth  investigating further. In the next sections we perform a general study of  one-loop models with type T3 and T1-iii topologies to determine their viability as minimal/accidental DM models. It will prove useful to consider models with the T3 topology first, then models with the T1-iii topology. The results of our study  have consequences for the models discussed thus far (and more generally).


\section{Models with the Type T3 One-Loop Topology\label{sec:T3}}
\begin{figure}[ttt]
\begin{center}
        \includegraphics[width = 0.50\textwidth]{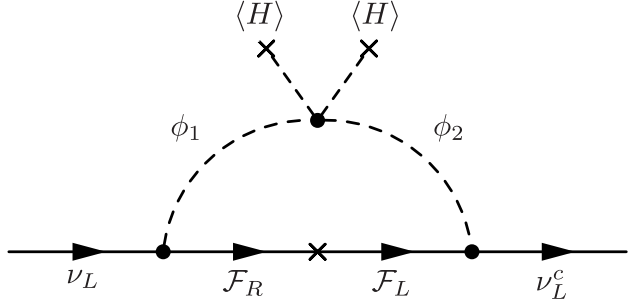}
\end{center}
\caption{The general one-loop diagram for neutrino mass via the type T3 topology.}\label{fig:general_T3}
\end{figure}

The general Feynman diagram for the type T3 one-loop topology  is shown in Figure~\ref{fig:general_T3}. This topology is well-known from  the Ma model~\cite{Ma:2006km} and related variants~\cite{Law:2013saa,Restrepo:2013aga}. Here we perform a general analysis of models with the type T3 one-loop topology, showing that they  do not preserve the accidental symmetry of the loop diagram. 

Models with the type T3 topology can be partitioned into two sets~\cite{Law:2013saa}, according to whether lepton number symmetry\footnote{We use the convention for lepton number symmetry in which only the fermions have nonzero charge. Conclusions do not depend on the chosen convention.}  is broken by a Yukawa coupling (corresponding to a complex intermediate fermion, $\f=\f_L+\f_R$), or by the mass insertion (corresponding to real  fermion, $\f_R=\f_L^c$). We first discuss the case of a complex intermediate fermion $\f$, such that $\f_L\ne\f_R^c$. The relevant Lagrangian  contains the following terms~\cite{Law:2013saa}:
\bea
\mathcal{L}&\supset& i\bar{\mathcal{F}}\gamma^\mu D_\mu\mathcal{F}\ -\  M_{\mathcal{F}}\;\overline{\mathcal{F}}\mathcal{F} \ +\ \sum_{i=1,2}\left\{|D^\mu \phi_i|^2 \ -\ M_i^2 |\phi_i|^2 \right\}\nonumber\\
& & +\ \lambda_1\; \overline{\mathcal{F}_R} L \phi_1^*\ +\ \lambda_2\; \overline{L^c}\mathcal{F}_L\phi_2\ +\ \lambda_{\phi H}\;\phi_1\phi_2^* H^2 +\mathrm{H.c.},\label{eq:L_T3_general}
\eea
where $L$ ($H$) is the SM lepton (scalar) doublet. Note the quartic term $\phi_1 \phi_2^* H^2$ automatically appears in the most-general Lagrangian, once the quantum numbers for $\phi_{1,2}$ are selected to permit the Yukawa terms  in Eq.~\eqref{eq:L_T3_general}, due to the identical quantum numbers of $L$ and $H$~\cite{McDonald:2013kca}. It is evident from Figure~\ref{fig:general_T3} that the loop diagram possesses an accidental symmetry with action $\{\f_{L,R},\,\phi_{1,2}\}\rightarrow\{-\f_{L,R},\,-\phi_{1,2}\}$. The key question is whether or not this accidental symmetry holds in the full Lagrangian, to give a minimal/accidental DM candidate. In what follows we show that  the most-general Lagrangian for the type T3 one-loop models  always contains additional terms that explicitly break the accidental symmetry.

The quantum numbers for the beyond SM fields are denoted as
\bea
\f\sim(1,R,Y),\quad \phi_1\sim (1,(R\pm1), -1-Y),\quad\phi_2\sim(1,(R\pm1),1-Y),
\eea
 where the choice of plus or minus for the $SU(2)$ quantum numbers of $\phi_1$ and $\phi_2$ can be made independently. We seek a model in which $\f$ and $\phi_{1,2}$ are charged under an accidental symmetry. The resulting  DM candidate should have vanishing hypercharge, to evade stringent direct-detection constraints.  However,  $\f$ is complex by construction, giving $Y\ne0$, so one must select  the value of $Y$ to ensure that one of the fields $\phi_{1,2}$ has vanishing hypercharge. This has two consequences: it restricts us to $Y=\pm1$, and also requires that $R\pm1$ is odd-valued, to ensure a neutral DM candidate,   giving even-valued $R$. To ensure there is no symmetry breaking cubic term $\phi^3$ for the hypercharge-less field in the scalar potential, it must have odd-valued $R\pm1=4n+3$, for $n=1,2,...$, so that $R\pm1\ge 7$, with even-valued $R\ge6$. For these values, the $SU(2)$ group product $R\otimes R$ contains the term $(R\pm1)\subset (R\otimes R)$. Thus, the following  Yukawa term is consistent with the electroweak symmetry and should appear in the most-general Lagrangian:
\bea
\mathcal{L}&\supset& \lambda_{\f\phi}\;\overline{\f_R} \; \f_L \;\phi_{1/2}\quad\quad \mathrm{for}\quad Y=\mp1.
\eea
This term explicitly breaks the accidental symmetry under which $\f$ and $\phi_{1,2}$ are charged, precluding  a minimal DM candidate for the T3 one-loop topology with complex intermediate fermions $\f$. 

As an example, consider the  type T3 model with particle content $\f_{L,R}\sim(1,6,-1)$, $\phi_1\sim(1,7,0)$ and $\phi_{2}\sim(1,5,2)$~\cite{Cai:2016jrl}, which employs the minimal DM septuplet. By the above reasoning,  the following Yukawa term is present:
\bea
\mathcal{L} &\supset& \lambda_{\f\phi} \; (\overline{F_R})^{abcmn}( \f_L)_{defmn}(\phi_1)_{abcd'e'f'}\epsilon^{dd'}\epsilon^{ee'}\epsilon^{ff'}.
\eea
This breaks the accidental symmetry, making the DM unstable. As a further blow to this model, the scalar potential also contains a cubic term that explicitly break the discrete symmetry, namely $\phi_1\phi_2^*\phi_2$. This term generalizes the terms previously identified in Ref.~\cite{Kumericki:2012bf}. 

More generally, related cubic terms always appear in the scalar potential for the T3 models with scalar DM; both $\phi_1$ and $\phi_2$ have odd-valued $SU(2)$ quantum numbers,\footnote{Recall that the choice of plus or minus is independent for $\phi_1$ and $\phi_2$, and that we are only considering models in which $R$ is large enough to ensure the model contains a minimal DM candidate.} $R\pm1$, and, denoting the hypercharge-less field as $\phi_i$, the most-general potential contains the term $\phi_i\phi_j^*\phi_j$, where $i\ne j$. Consequently one can also exclude models with type T3 one-loop topology (with complex intermediate fermions) as minimal DM frameworks due to explicit symmetry breaking in the scalar potential.

\begin{figure}[ttt]
\begin{center}
        \includegraphics[width = 0.50\textwidth]{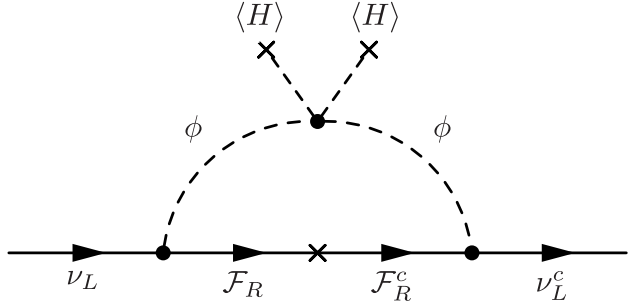}
\end{center}
\caption{The type T3  one-loop diagram for neutrino mass  in the case with a real intermediate fermion $\f_R\sim(1,R,0)$.}\label{fig:realfermion_T3}
\end{figure}

Now let us turn our attention to the type T3  one-loop  models with real fermion $\f_R\sim(1,R,0)$, which must have odd-valued $R$. The Feynman diagram  takes the particular form in Figure~\ref{fig:realfermion_T3} where the exotic scalar has the quantum numbers $\phi\sim(R\pm1,-1)$. The Lagrangian contains the following terms, which are relevant for the mass diagram
\bea
\mathcal{L}&\supset& i\bar{\mathcal{F}_R}\gamma^\mu D_\mu\mathcal{F}_R - \frac{M_{\mathcal{F}}}{2}\;\overline{\mathcal{F}_R^c}\mathcal{F}_R +|D^\mu \phi|^2 -M_\phi^2 |\phi|^2 +\lambda\overline{\mathcal{F}_R}\,L \phi^*+\lambda_{\phi H}(\phi H)^2. \label{eq:realfermion_lagrangian}
\eea
These terms admit the accidental symmetry $\{\f_R,\,\phi\}\rightarrow\{-\f_{R},\,-\phi\}$. However, one can prove that the model always allows the quartic term $\sim H\phi^*\phi^2$, which explicitly breaks the discrete symmetry. The precise contraction for the $SU(2)$ indices depends on whether $(R\pm1)/2$ is odd- or even-valued. For even-valued $(R\pm1)/2$, one can contract the doublet $H$ onto the conjugate field $\phi^*$, giving $H_a(\phi^*)^{a\ldots}$, while for odd-valued $(R\pm1)/2$ one may contract the SM doublet with the non-conjugated field $\phi$, giving $\epsilon^{aa'}H_a \phi_{a'\ldots}$. In either case, the accidental $Z_2$ symmetry is explicitly broken.

To give some  examples, for $R\pm1=4$, one has even-valued $(R\pm1)/2=2$ and the contraction of $SU(2)$ indices for the doublet  is $H_a(\phi^*)^{a\ldots}$, giving
\bea
H_a(\phi^*)^{abc}\;\phi_{bmn}\phi_{cm'n'}\;\epsilon^{mm'}\epsilon^{nn'}.
\eea
As shown in Ref.~\cite{Kumericki:2012bf}, this term breaks the $Z_2$ symmetry advocated in Ref.~\cite{Cai:2011qr}. Similarly, a model with $R\pm1=8$, contains the term
\bea
H_a(\phi^*)^{abcdefg}\;\phi_{bcdmnop}\phi_{efgm'n'o'p'}\;\epsilon^{mm'}\epsilon^{nn'}\epsilon^{oo'}\epsilon^{pp'},
\eea
which extends the results of Ref.~\cite{Kumericki:2012bf}. The generalization for larger even-valued $(R\pm1)/2$ is evident: for $(R\pm1)/2=2N$, with $N=1,2,...$, one has $2N$ occurrences of the $\epsilon$-tensor contracting $\phi^2$, avoiding an anti-symmetric contraction of this symmetric factor.

On the other hand, for  $R\pm1=6$ one has odd-valued $(R\pm1)/2$=3, and the contraction for the doublet is $\epsilon^{aa'}H_a \phi_{a'\ldots}$, giving
\bea
\epsilon^{aa'}H_a(\phi^*)^{bcdef}\phi_{a'bcmn}\phi_{defgm'n'}\;\epsilon^{mm'}\epsilon^{nn'}.
\eea
This also generalizes in an obvious way for larger odd-valued $(R\pm1)/2$: with $(R\pm1)/2=2N+1$, one has $2N$ $\epsilon$-tensors contracting $\phi^2$. The key point is that for both odd-valued and even-valued $(R\pm1)/2$, one has an even number of $\epsilon$-tensors contracting $\phi^2$, ensuring the contraction is symmetric and the term $H\phi^*\phi^2$ is consistent with the gauge symmetry. Thus, in all cases the accidental discrete symmetry of the neutrino mass diagram is broken by the scalar potential, rendering the DM unstable.

To summarize, we have shown that models in which the SM is extended by a minimal set of exotic fields, such that neutrinos acquire mass via  the type T3 one-loop diagram, do not contain viable minimal DM candidates. This is because the accidental symmetry, apparent in the loop diagram, is explicitly broken by other couplings in the model. For the case of a complex intermediate fermion, $\f=\f_L+\f_R$, the model contains a Yukawa coupling  between $\f$ and either $\phi_{1}$ or $\phi_2$, as well as a symmetry breaking cubic term in the scalar potential. In the alternative case with a real fermion $\f_R$, the most-general scalar potential always contains a term $H\phi^*\phi^2$. In both cases the offending terms contain three $Z_2$-odd fields and explicitly break the discrete symmetry.

\section{Models with the Type T1-iii One-Loop Topology\label{sec:T1}}

Next we undertake a more general study of models with type T1-iii one-loop topology. As shown in Section~\ref{sec:scalar_T1-iii}, type T1-iii models with the scalar $\phi\sim(1,5,0)$ contain a cubic term in the scalar potential that explicitly breaks the accidental $Z_2$ symmetry of the loop diagram. This problem is cured by modifying the field content of e.g.~Model B, giving the new Models I and III outlined above. In this section we determine if the type T1-iii models contain any other terms that break the accidental symmetry.

For the present discussion, we relabel the fields in the type T1-iii models as 
\bea
\chi\rightarrow \f_L\sim(1,R,0), \quad\mathrm{ and }\quad \bar{\psi}\rightarrow \psi_R\sim(1,R\pm1,-1),
\eea
with the hypercharge -1 choice evident from explication on Figure~\ref{fig:one_loop}, and with
 odd-valued R. Here, we first consider the case with a real intermediate fermion. The scalar then has the quantum numbers 
\bea
\phi\sim(1, R_\phi,0), \quad\mathrm{with}\quad R_\phi\in\{R,\;R\pm2\},
\eea 
where the choice for $SU(2)$ quantum numbers for $\phi$ and $\psi_R$ can be made independently. Note that these three fields ($\f_L$, $\psi_R$ and $\phi$) comprise a sufficient set to allow the one-loop diagram with type T1-iii topology. The resulting loop diagram admits an accidental symmetry with action $\{\f_L,\,\psi_R,\,\phi\}\rightarrow\{-\f_{L},\,-\psi_R,\,-\phi\}$.

In Section~\ref{sec:scalar_T1-iii} we discussed criterion under which the accidental symmetry of the loop diagram is shared by the scalar potential,  the key point being that  the quantum numbers for $\phi$ must preclude a cubic term $\phi^3$ in the potential. Here, we turn our attention  to the Yukawa sector. The main concern for these models apparently comes from the term
\bea
\mathcal{L}&\supset& \lambda_{\f \phi}\,\f_L^T \mathcal{C}^{-1} \,\f_L \phi,\label{eq:generalFphi}
\eea
which appears to be allowed (here $\mathcal{C}$ is the charge-conjugation matrix). Using the explicit example of Model I, where $\f_L\sim(1,5,0)$ and $\phi\sim(1,7,0)$, the contraction of $SU(2)$ indices is\footnote{One can write this with $\phi$ rather than $\phi^*$, but then more $\epsilon$-tensors appear to clutter the expression.} 
\bea
\mathcal{L}&\supset& \lambda_{\f \phi}\,(\f_L^T)_{abcd} \,\mathcal{C}^{-1} \,(\f_L)_{efgh} (\phi^*)^{bcdfgh}\,\epsilon^{ae}.\label{eq:yukawaFT3}
\eea
However, with regards to Lorentz symmetry, the standard Majorana contraction is symmetric under interchange of the two fields. Thus, taking the transpose and interchanging dummy labels, one can show that this term vanishes identically\footnote{With more than one generation of $\f$, one would be able to include this term, provided the couplings are taken anti-symmetric in generation space.} due to the antisymmetric contraction of the symmetric product of two quintuplet fermions $\f_L$.  Note, however, that if one instead used  $\phi\sim(1,5,0)$, as in  Model B, this term has the $SU(2)$ contraction
\bea
\mathcal{L}&\supset& \lambda_{\f \phi}\,(\f_L^T)_{abcd} \,\mathcal{C}^{-1} \,(\f_L)_{efgh} (\phi^*)^{cdgh}\,\epsilon^{ae}\epsilon^{bf}.\label{eq:yukawaFT3'}
\eea
Now the symmetric Majorana product of two $\f_L$'s is contracted by an even number of $\epsilon $ factors (two in this case), so the term is allowed. Thus, Model B contains an additional source of accidental symmetry breaking in the Yukawa sector. More generally, one must check the particular quantum numbers of $\f_L$ and $\phi$, for a given model, to determine if this Yukawa term is present. 

For Models I, II and III identified in Eq.~\eqref{eq:modelsI}, the Yukawa term in Eq.~\eqref{eq:generalFphi} vanishes. Consequently one finds that the full Yukawa sector of these models (obtained by adding $\psi_R$, $\f_L$ and $\phi$ to the SM), preserves the accidental symmetry of the loop diagram. This result is appealing and seemingly indicates that Models I, II and III contain good minimal DM candidates. However, theories comprised of  the SM plus $\f_L$, $\psi_R$ and $\phi$, are inconsistent due to the masslessness of $\psi_R$ and the presence of quantum anomalies.\footnote{Strictly speaking,  $\psi_R$ has no bare mass but does acquire mass after electroweak symmetry breaking. However, this mass is too small, being lighter than the $\mathcal{O}(10)$~TeV DM mass for perturbative  parameter values, so an additional source of mass is required.} The solution employed in Ref.~\cite{Cai:2016jrl} was to add an additional field $\psi_L\sim(1, R\pm1,-1)$ such that $\psi_{L,R}$ form a Dirac pair. This allows a bare mass for $\psi=\psi_L+\psi_R$ and removes the anomaly. 

Technically speaking, this approach, which is inherited by Models I and III, employs more fields than are required to generate the neutrino loop diagram. None the less, the resulting spectrum appears to give a minimal construct realizing type T1-iii models. Evidently, the addition of $\psi_L$ to the spectrum does not modify the scalar potential of the model, which retains the accidental symmetry of the loop diagram. However, one must reconsider the Yukawa sector in the presence of $\psi_L$, to determine if the accidental symmetry remains viable.

By construction, the models now contain the mass term $M_\psi \overline{\psi_R}\psi_L+\mathrm{H.c.}$ However, the $SU(2)$ product $\overline{\psi_R}\otimes \psi_L$ contains additional terms and, in particular, contains the term $R_\phi \subset R_\psi\otimes R_\psi$, where $R_\psi=R\pm1$.   Thus, the type T1-iii models with $\psi_L$ in the spectrum contain the Yukawa term
\bea
\mathcal{L}&\supset& \lambda_{\psi\phi}\, \overline{\psi_R}\,\psi_L\, \phi+\mathrm{H.c.},
\eea
which explicitly breaks the accidental $Z_2$ symmetry. To give an explicit example, for Model A, with $\phi\sim(1,7,0)$ and $\psi\sim(1,6,-1)$, the Lagrangian contains the terms 
\bea
\mathcal{L}&\supset& \lambda_{\psi\phi}\, (\overline{\psi_R})^{abcde}\,(\psi_L)_{abfgh}\, \phi_{cdef'g'h'}\epsilon^{ff'}\epsilon^{gg'}\epsilon^{hh'}+M_\psi (\overline{\psi_R})^{abcde}\,(\psi_L)_{abcde},
\eea
where we include the mass term for completeness. This Yukawa  term explicitly breaks the accidental symmetry of the loop diagram, ultimately rendering the DM unstable, and making Model A unsuitable as a framework for minimal DM. Model B contains a similar term, and thus the accidental symmetry of the loop diagram is broken explicitly by both the Yukawa Lagrangian and the scalar potential in this case. More generally, the Yukawa term is automatically present and is inherited by all of the Models I, II and III, meaning these, and related models, fail to admit an accidental symmetry. We conclude that  type T1-iii one-loop models with real fermion $\f_L$ do not provide minimal DM candidates. 

Note that the Yukawa difficulties in  the T1-iii models stem from the use of the field $\psi_L$ to give mass to $\psi_R$ and cure the quantum anomalies. One could ask if the accidental symmetry could be retained by instead employing a more extended sector to cure these problems, rather than the minimal choice of $\psi_L$. Such a model quickly becomes complicated - if $\psi_L$ has different quantum numbers to $\psi_R$, then even more fields are needed to remove the anomaly, along with a new scalar (that contributes to electroweak symmetry breaking) to give mass to $\psi_R$. Such models go against the spirit of the minimal DM framework, requiring multiple fields beyond the minimal content required to generate radiative neutrino mass and give accidental DM. We do not explore this possibility further.

\begin{figure}[ttt]
\begin{center}
        \includegraphics[width = 0.50\textwidth]{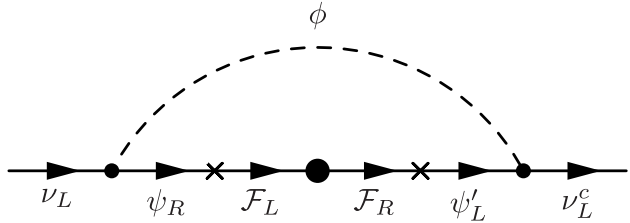}
\end{center}
\caption{General one-loop diagram for neutrino mass with the type T1-iii topology. Crosses denote insertions of the SM Higgs vacuum value and the larger dot denotes a  mass insertion for the complex fermion $\f$.}\label{fig:T1_iii_Dirac}
\end{figure}

Next, we turn our attention to type T1-iii models with complex fermion $\f_{L,R}\sim(1,R,Y)$, where $Y\ne0$. In this case, one has $\chi\rightarrow \f_L$ and $\chi^c\rightarrow \f_R$, while also allowing $\psi^c\rightarrow \psi_L'$ to have different quantum numbers to $\psi_R$, as shown in Figure~\ref{fig:T1_iii_Dirac}. In such a model, one must ensure that both $\psi_R$ and $\psi_L'$ possess non-vanishing hypercharge - if either of these fields  has $Y=0$,  the model admits a Majorana mass for that field, and combined with the coupling to $\phi$, the model automatically generates a one-loop diagram with the type T3 topology.\footnote{The presence of a $(\phi H)^2$ term follows from the identical quantum numbers of $H$ and $L$.} We have already shown that the particle content of T3 models admits explicit symmetry breaking terms, so this case should be avoided. Combining this demand with the fact that complex $\f$ has $Y\ne0$, by construction, one can show that the scalar $\phi$ also has non-vanishing hypercharge. Thus, the minimal field content does not include any multiplets that contain viable minimal DM candidates, due to the non-vanishing hypercharge of the exotics. One could consider extending the models to include additional fields (possibly also needed to generate mass for the fermions and avoid anomalies), one of which may give a DM candidate. We do not consider this here, as it is contrary to the minimal DM approach. Regardless, we can conclude that minimal radiative models of the type T1-iii topology, with either complex or real fermions $\f$, do not give viable accidental DM candidates.

\section{Conclusion\label{sec:conc}}

Motivated by a recent study of one-loop models for neutrino mass with minimal DM~\cite{Cai:2016jrl}, we have undertaken a general analysis of one-loop models with DM candidates. We showed that Model B, which employs the T1-iii topology, contains an explicit $Z_2$ symmetry breaking term in the scalar potential as well as a symmetry breaking Yukawa coupling. We demonstrated that these short-comings could be cured by modifying the particle content, seemingly realizing two candidate T1-iii models for minimal DM. However, we further showed that \emph{all} of the proposed models contain an additional source of explicit $Z_2$ symmetry breaking in the Yukawa Lagrangian. Thus, \emph{none} of the proposed models, with either the type T3 or type T1-iii topologies, provide viable minimal DM candidates, with some models containing multiple sources of explicit $Z_2$ symmetry breaking.

Our general study of the type T3 and T1-iii topologies, which included both real and complex intermediate fermions, showed that neither of these topologies gives viable minimal DM candidates. Combined with the failure of the T1-i and T1-ii topologies~\cite{Cai:2016jrl}, our results indicate that none of the irreducible one-loop topologies are expected to give viable minimal/accidental DM candidates. Thus, at present, the only known viable model of radiative neutrino mass that does not enforce any beyond-SM symmetries, and yet gives a stable DM candidate due to an accidental symmetry, remains as the three-loop model of Ref.~\cite{Ahriche:2015wha}. Similarly, the only known models that employ both the fermion quintuplet and scalar septuplet DM candidates of Ref.~\cite{Cirelli:2005uq} are the three-loop model of Ref.~\cite{Culjak:2015qja} and the new Model~I defined in  Eq. (5) (with symmetry imposed).  As a direction for further study, it would be interesting to determine if two-loop models of radiative neutrino mass can give viable minimal/accidental DM candidates. 

Before concluding we note that a general discussion of non-renormalizable operators in minimal DM models appears in Ref.~\cite{DelNobile:2015bqo}. The approach of radiative neutrino mass models with minimal DM is perhaps more agnostic than Ref.~\cite{Cirelli:2005uq} regarding the details of the UV completion and the impact of non-renormalizable operators - the presence of multiple large multiplets, required to generate neutrino mass, generally causes a Landau pole below the Planck scale. Thus, one must typically assume that the details of the UV completion preserve the accidental symmetry of the renormalizable Lagrangian to sufficient accuracy,  much as one assumes new TeV-scale  physics would preserve the accidental baryon symmetry of the SM to sufficient accuracy to ensure proton longevity.

Note Added: A revised version of Ref.~\cite{Cai:2016jrl} appeared after this work was completed,
also noting that one-loop models do not appear to work. Importantly, it emphasizes that the phenomenology of the T3 and T1-iii models can remain interesting if a discrete symmetry is imposed (even approximately). In this regard, it could be interesting to contrast the phenomenology of e.g.~Model I with the results of Ref.~\cite{Cai:2016jrl}.

\section*{Acknowledgments\label{sec:ackn}}
AA is supported by the Algerian Ministry of Higher Education and
Scientific Research under the CNEPRU Project No.~D01720130042. IP is supported by the Croatian Science Foundation under the project number 8799. KM is supported by the Australian Research Council and acknowledges a 2013 correspondence with J.~Heeck.

\appendix


\end{document}